\documentclass[a4paper,10pt]{article}

\begin{document}

\title{Is the apparent acceleration of the Universe expansion driven by a dark energy-like component or by inhomogeneities?}
 
\author{{\bf Marie-No\"elle C\'el\'erier} 
\\
Laboratoire Univers et TH\'eories (LUTH)  \\
Observatoire de Paris-Meudon \\
5 place Jules Janssen, Meudon, 92195 cedex, France \\
E-mail: marie-noelle.celerier@obspm.fr}

\maketitle

\begin{abstract}

Since its decovery during the late 90's, the dimming of distant SN Ia apparent luminosity has been mostly ascribed to the influence of a mysterious dark energy component. Based upon the cosmological ``principle'' hypothesis, this interpretation has given rise to the ``concordance'' model, developed in the context of a  Friedmann-Lema\^itre cosmology. However, a caveat of this reasoning is that the cosmological ``principle'' derives from a philosophical Copernican assumption and has never been tested. Furthermore, a weakness of its conclusion, i. e., the existence of a negative-pressure fluid or a cosmological constant, is that it would have profound implications for the current theories of physics. This is why we have proposed a more conservative explanation, ascribing the departure of the observed universe from an Einstein-de Sitter model to the influence of inhomogeneities. This idea has been independently developed by other authors and further enlarged to the reproduction of different cosmological data. We review here the main proposals which has been made along these lines of though and present some prospects for future developments.

\end{abstract}

\section{Introduction}\label{sec1}

Since its decovery during the late 90's, the dimming of distant SN Ia apparent luminosity has been mostly ascribed to the influence of a mysterious dark energy component \cite{PR03} which is supposed to drive an acceleration of the Universe expansion. 

Based upon the cosmological ``principle'' hypothesis and coupled to an analysis of the other cosmological data, this interpretation has given rise to the ``concordance'' or $\Lambda$CDM model, developed in the context of a Friedmann-Lema\^itre cosmology. A drawback to this reasoning is that the cosmological ``principle'' merely derives from a philosophical Copernican assumption which still deserve to be tested \cite{MNC00,MNC05}.

The dark energy component, a cosmological constant or a negative pressure fluid, represents, in the ``concordance'' model, about 73\% of the Universe energy density. However, a cosmological constant is usually interpreted as a vacuum energy of which particle physics cannot explain such an amplitude and a negative pressure fluid remains a mysterious phenomenum. This is known as the cosmological constant problem.

Another feature of the supernova data analysed in a Friedmannian framework is to yield a late-time acceleration of the expansion rate, about the epoch when structure formation enters the nonlinear regime. This would imply that we live at a time when matter density energy and dark energy are of the same order of magnitude. This is known as the coincidence problem. 

General relativity having only been tested up to scales of order a planetary system, a second type of explanation has been proposed which implies a modification of this theory at larger distance scales \cite{C04C06}.

Our current purpose is to review the works dedicated to a more simple and natural proposal, which makes only use of known physics and phenomena. Since it appears that the onset of the apparent acceleration and the beginning of structure formation in the Universe are concomitant, the idea that the SN Ia observations could be reproduced by the effect of inhomogeneities have been put forward. This interpretation has been first proposed independently by a few authors \cite{MNC00,DH98,PS99,KT00}, shortly after the release of the data. Then, after a period of relative disaffection, it has experienced a reniewed interest, specially during the last two years.

Another natural explanation is the possible effect of an actual geometrical cosmological constant. Its value such as it appears in the $\Lambda$CDM model has been indeed predicted in the framework of scale relativity, as $\Lambda = 1.36$ $10^{-56} cm^{-2}$, long before the release of the supernova results \cite{LN93,LN96}. However, we shall not develop this approach here since it is beyond the scope of this contribution.

\section{Presentation and discussion of the methods} \label{meth}

It is not obvious that one can keep using FLRW models to interpret the high precision observational data collected in a regime where most of the mass is clumped into or is forming structures. Three main physical effects might actually be missing \cite{BM06}: (i) the overall (average) dynamics of such a universe could be significantly different from the FLRW one. This effect is usually taken into account by calculating so-called backreaction terms added to effective equations for the dynamical evolution of the physical quantities under study, (ii) light propagation in a clumpy universe might be different from that in a homogeneous one and the luminosity distance-redshift (LDR) relation could hence be affected, (iii) the fact that we have only one observer could influence the results, since we might live in an under or over-dense region which should induce significant corrections in the data interpretation.

What is the most subtle point of this issue is that what is observed in the supernova data is {\it not} an accelerated universe expansion (this is only an artefact due to the a priori assumption that our universe, even in the local region of structure formation, can be represented by a FLRW model) but only a dimming of the SN Ia luminosity taken into account by the infered LDR relation.

However, many authors have focussed their attention on effect (i) and backreaction has been studied with two types of methods: (a) in the linear regime of small amplitude fluctuations, perturbative expansions which were subsequently averaged out, (b) for the more general problem of the dynamics induced by all scale inhomogeneities including the nonlinear regime, spatial averaging of nonperturbed models. The local dynamics has also been analysed using different exact solutions of the general relativity equations.

Effects (ii) and (iii) have been mostly studied using toy models constructed with such a kind of exact nonaveraged inhomogeneous solutions. 

\subsection{Perturbative analysis} \label{pa}

The perturbative analysis is a method employed when the deviations from homogeneity and isotropy are assumed to be ``small'' (see Mukhanov et al. \cite{MFB92} for a review).

One considers two spacetimes, the physical, perturbed spacetime and a fictious background spacetime described by a FLRW model. A one-to-one correspondence between points in the background and points in the physical spacetime carries the coordinates labelling the points in the background over into the physical spacetime and defines a choice of gauge. The perturbation in some quantity is the difference betwen the values it possesses at a point in the physical spacetime and at the  corresponding point in the background spacetime. Since we are interested in considering the influence of inhomogeneities, the perturbed quantity is here the density.

To study the influence of such fluctuations on the expansion rate of the Universe, one identifies a local physical variable which describes this expansion rate, calculates the backreaction of the cosmological perturbations on this variable to a given order, and {\it then} spatially averages the result. Note that it is of the upmost importance to avoid the deficient procedure consisting of calculating an ``observable'' from the spatially averaged metric, which, in general, does not give the same result as calculating the spatial average of the observable \cite{WU98,GB02}. As we shall stress in Sec. \ref{spataver}, it is also crucial to define the hypersurfaces on which the averaging is performed by a clear physical prescription.

The cleanest available calculation of the effect of density fluctuations on the averaged expansion rate of a matter-dominated universe up to second order in the metric variables has been performed by Kolb et al. \cite{KM05a} for an application to superhorizon scale perturbations, in the adiabatic case. Owing to the perturbation development employed, this method is only consistent for the study of the effect of very large scale fluctuations, with small amplitude, e. g., in the linear regime of structure formation \cite{KM05a,BM05,BD06}. It is also sensitive to the order at which the development is performed. It has been demonstrated by Notari \cite{AN05} that, while at early times the contribution of subhorizon inhomogeneous gravitational fields is perturbatively subdominant, the series is likely to diverge around the time of structure formation. There one must employ a nonperturbative method. 

To deal with this issue either spatial averaging of a nonperturbed solution or exact solutions of Einstein's equations are usually used. A new formalism for the study of nonlinear perturbations in cosmology, based on a covariant and fully nonperturbative approach, has recently been proposed \cite{LV06} but it has not yet been applied to our purpose.

\subsection{Spatial averaging of nonperturbed models} \label{spataver}

Spatial averaging aims at obtaining the impact of a given inhomogeneity profile upon the assumed large scale FLRW background, in terms of backreaction terms added to the Friedmannian evolution equations of scalar quantities such as the expansion rate, the energy density, the isotropic pressure. These backreaction terms can be assimilated to a black energy component if they exhibit the right properties. 

In general relativity, spatial averaging is very much involved \cite{AK97} since the equations which determine the metric tensor and the quantities calculated from it are highly nonlinear. However, when modelling the Universe, the usual method is to use continuous functions representing, e. g., energy-density, pressure, or other kinematical scalars of the velocity field, implicitly assuming that they represent volume averages of the corresponding fine-scale quantities. But we know that our local Universe is highly inhomogeneous from the scales of planetary systems up to currently unknown distances \cite{SL98}. Anyhow, the scale, i. e., the size of the volume, over which the averagings are performed are never explicitly defined, while the results of this process obviously depend on it.

Moreover, a volume average is a well-defined quantity for scalars only. For vectors, and all the more for tensors, it leads usually to noncovariant quantities. Another drawback is that a gauge problem arises when relating the ``true'' and the averaged metric. Scalar quantities only are invariant under coordinate transformations, not tensors. One must also be aware that, in a generic spacetime, there are no preferred time-slice one could average over and the results depend on the choice of the hypersurfaces on which this average is performed \cite{SR04a}.

But the main issue is the noncommutating property of the two operations: averaging the metric and calculating the Einstein tensor. In other words, the Einstein tensor calculated from an averaged metric and energy-momentum tensor is not equal to the Einstein tensor first calculated from the fine-scale metric and energy-momentum tensor, then averaged. However, in the standard cosmological approach, the Universe is modelled by adopting exactly the wrong method: take a metric which is assumed already averaged, calculate the corresponding Einstein tensor and equate it to an already averaged energy-momentum tensor. 

The question raised by the method consisting of determining the parameters of an a priori assumed FLRW model from observational data is the ``fitting problem'' \cite{ES87}. The departure of the ``real'' Universe from the averaged one is known as the ``backreaction'' effect. The study of backreaction can be implemented by two methods: either one tries to obtain directly the equations satisfied by the averaged quantities, with minimum assumptions as regards an underlying background, or one takes as a background a FLRW model and analyses the effect of linear perturbations applied on it. In this section, we consider the first approach. The second one has already been studied in Sec. \ref{pa}.

This issue has been delt with by Buchert and collaborators for simplified inhomogeneous cosmological models with an irrotational perfect fluid as the gravitational source. First, Buchert and Ehlers \cite{BE97} have proposed a simple averaging procedure which have been used by Palle \cite{DP02} to deal with the cosmological constant problem. Then, Buchert \cite{TB00,TB01} has developed another procedure aimed at constructing an ``effective dynamics'' of spatial portions of the Universe from which observable average characteristics can be infered like the Hubble constant, the effective 3-Ricci scalar curvature and the mean density (and isotropic pressure) of a spatial domain.

Relations between average scalar sources (energy density, pressure) and an average scalar geometric quantity, the expansion rate, have been derived for an adapted foliation of spacetime. They involve domain dependent backreaction terms that have been splitted into a ``kinematical'' backreaction comparing the variance of the expansion rate to the shear, and a ``dynamical'' backreaction, i. e., pressure forces. These equations show that the averaged shear fluctuations tend to increase the expansion rate as do the averaged energy source terms (provided the energy condition holds), while the averaged expansion fluctuations have an opposite effect and therefore work to a stabilization of structures. The dynamical backreaction can do both.

However, an averaging procedure is not complete unless one also averages the {\it geometrical} inhomogeneities. Since geometrical fields are tensorial variables for which possible strategies of averaging are not straightforward, Buchert and Carfora \cite{BC02} have suggested a Lagrangian smoothing of these variables as opposed to their Eulerian averaging on spatial domains. Curvature fluctuations turn out to be crucial and may even outperform the effect of kinematical fluctuations. The authors define effective cosmological parameters that would be assigned to the smoothed cosmological spacetime. These parameters are ``dressed'' after smoothing out the geometrical fluctuations. Relations between the ``dressed'' and ``bare'' parameters are derived. The former provide the framework for interpreting observations with a ``Friedmann bias'', i. e., as if the observer was living in a Friedmannian universe. The latter represent the actual inhomogeneous cosmological model, spatially averaged.

In subsequent articles \cite{BC03a,BC03b}, the same authors have identified two effects that quantify the difference between ``bare'' and ``dressed'' parameters: the ``curvature backreaction'' and the ``volume effect''. To summarize we can say that, in the smoothed model, the averaged scalar curvature is dressed both by the volume effect and the curvature backreaction effect. The volume effect is expected when comparing two regions of distinct volumes, but with the same matter content, in a constant curvature space. The backreaction term encodes the deviation of the averaged scalar curvature from a constant curvature model, e. g., a FLRW space section.

However, the interpretation of cosmological parameters remains far from trivial. One must also average on the observer's light cone in which case the above effects interact with the time evolution of the model. Moreover, smoothing will have to be performed in a dynamical setting. Apostolopoulos et al. \cite{PA06} have uncovered another subtelty of inhomogeneous cosmology: the volume increase at a given point results from averaging over various directions a possible anisotropic acceleration parameter and might thus not be the most appropriate for an expansion characterization.

\subsection{Use of exact solutions of Einstein's equations} \label{exsol}

The method consisting in using exact solutions to modelize the inhomogeneities observed in the Universe is the most straightforward and devoided of theoretical pitfalls. It is adapted to represent either strong or weak inhomogeneities.

For mathematical simplification and also to account for the local quasi-isotropy of the CMB as measured on our wordline, most of the retained models exhibit spatial spherical symmetry. Therefore, some authors have claimed that unphysical properties of this symmetry, and specially of the  Lema\^itre-Tolman-Bondi (LTB) class \cite{KB05,VF06}, might prevent these models to be used in this framework, specially when the observer is assumed to be located at the centre. However, in the literature, the observer has been put either at the center of the model \cite{GS04,AA06} or offcenter \cite{JM05}. Moreover, a spherically symmetric model can represent an anisotropic inhomogeneous universe averaged over angular scales, which is not physically worse than a uniformly averaged universe such as in the FLRW picture.

Other spherically symmetrical models has been used in this framework, i. e., peculiar classes of Stephani solutions. One feature of the Stephani models that has been the subject of much debate is their matter content \cite{LP86,AK97,BC00}. Individual fluid elements can behave in a rather exotic manner, e. g., exhibiting negative pressure. But this is also the case of dark energy.

Avoiding spherical symmetry, Moffat \cite{JM06a} has considered a peculiar class of models of the Szafron family. Other exact solutions have consisted of FLRW patch(es) embedded in a FLRW background with different energy densities \cite{KT00,KT01,KT03}.

Depending on the authors, these exact solutions have been used either to barely fit the data as they are obtained by observation or to provide the inhomogenous models on which an averaging procedure ``a la Buchert'' is implemented \cite{SR06a}.

\section{Studied physical quantities} \label{physqu}

\subsection{The deceleration parameter} \label{decpar}

When reasoning in the framework of a Friedmannian cosmology, the dimming of the supernovae is associated with an acceleration of the Universe expansion. This is why a number of authors have focussed on the issue of either demonstrating or ruling out an effect of the inhomogeneities on the expansion rate.

Some have tried to derive \cite{HS05,EF05} or rule out \cite{PA06, JM06b} no-go theorems. However, when spatially averaged, a physical quantity associated with the expansion rate behaves quite differently \cite{JM06b,JM06c,KK06}. It is thus difficult to yield general rules from such theorems.

Some have stressed that the definition of a deceleration parameter in an inhomogeneous framework is tricky. Hirata and Seljak \cite{HS05} have proposed four different definitions of such a parameter. Apostolopoulos et al. \cite{PA06} have shown that an observer located away from the centre of a spherically symmetric configuration can measure acceleration or deceleration along radial or perpendicular directions depending on local under or over-density. This demonstrates that, even locally, the effect of inhomogeneities on the dynamics of the Universe is not trivial.
  
Ishibashi and Wald \cite{IW06} have also argued that an averaged quantity representing the scale factor or the deceleration parameter may accelerate without there being any observable consequence.

Another unexpected effect has been put forward by Tomita \cite{KT00,KT01,KT03} who has considered a cosmological model composed of a low-density inner homogeneous region connected to an outer homogeneous region of higher-density. Both regions decelerate, but, since the void expands faster than the outer region, an apparent acceleration is experienced by the observer located inside the void.

We can therefore conclude that the computation of some local quantity (generally the deceleration parameter), eventually subsequently averaged, and behaving the same way as in FLRW models with dark energy can lead to spurious results \cite{GB02,SR04b,BC06} and must therefore be avoided.

\subsection{The Effective Stress Energy Tensor (SET)} \label{SET}

Martineau and Brandenberger \cite{MB05} have tried to estimate the effect of a backreaction of Super-Hubble modes, by computing it in terms of the second order Effective SET of cosmological perturbation theory. This has been criticized by Ishibashi and Wald \cite{IW06} who have argued that a large SET implies the contribution of higher order terms in the perturbative scheme, therefore impairing the claimed results.

\subsection{The LDR relation} \label{ldr}

The LDR relation is the only direct product of the supernova data, obtained without any a priori cosmological assumption. This relation is therefore the best observable to be fitted and this has been the aim of a number of works \cite{MNC00,PS99,BM06,PA06,KB05,GS04,AA06,BC00,IN02,DG06,DH98,SJ01}.

\section{Main physical results} \label{physres}

It has been proposed that energy density perturbations of wave-length larger than the Hubble radius, generated during inflation, might produce upon cosmic parameters effects that could mimic an accelerated expansion of the Universe \cite{KM05a,KM05b,BM05,MB05}. But it has subsequently been shown by a number of authors \cite{HS05,EF05,GC05,TB05,SR06b}, among whom some of the proponents themselves \cite{KM05c,KM05d}, that this could not be the case. 

Working in the framework of a peculiar toy model, Biswas et al. \cite{BM06} have computed a minimal overall (average) effect which amounts to a correction in apparent magnitudes at all redshifts of order $\Delta m \sim 0.15$.

This effect is to be added to the estimation made by Buchert and Carfora \cite{BC03b} of the volume effect alone in a naive swiss-cheese model. Its magnitude of 67\% reduces the necessary dark energy in the concordance model roughly from 70\% to 50\% \cite{TB06}. A mismatch of similar magnitude has been reported by Hellaby \cite{CH88}, using volume matching \cite{ES87} in LTB models of clusters and voids.

Studies of the constraints which apply to a dust model of universe if one wants to explain dark energy by the backreaction effect of inhomogeneities have shown that a negative average spatial curvature is mandatory to compensate a strong backreaction increasing the expansion rate of the model \cite{TB05,TB06,SR04a,BL06, PS06,JM06b,JM06c}. This result is actually consistent with the fact that most of the exact inhomogeneous models which reproduce best the observed LDR relation are those which exhibit negative spatial curvature near the observer, or, equivalently, where the observer is located in an underdense region \cite{BM06,AA06,KT01,DG06,SJ01,RM05}. The assumption that we might be located within an underdense region seems to be consistent with observations leading to the identification of a Local Void and of its suggested expansion \cite{KT03,IO04}.

The matching of the observed LDR relation has been successfully performed by a number of proposed models: homogeneous void models \cite{KT01}, LTB models with a centered observer \cite{AA06,IN02,DG06}, LTB models with an outcentre observer \cite{BM06,RM05}, Stephani models \cite{BC00,DH98,SJ01,GS04}. Most of these models reproduce the $\Lambda$CDM LDR relation up to a redshift of the order unity. This feature, which has been considered as a ruling out drawback by some authors \cite{KB05,VF06}, must be viewed as a nice way out of the coincidence problem. The physical explanation is that the appearance of small scale inhomogeneities corresponds to the onset of structure formation at the time (around $z \sim 1$) where apparent ``acceleration'' begins to be observed \cite{SR06a}.

However, it has been shown \cite{MNC00,MB98} that the problem of fitting a given LDR relation with a peculiar inhomogeneous (LTB) model is completly degenerate. Therefore, the model parameters must be constraint by other cosmological data. This has been done by some authors: the model proposed by Biswas et al.\cite{BM06} reproduces matter abundance, the first acoustic peak in the CMB power spectrum and the baryon oscillations; Alnes et al. \cite{AA06} obtain the matter density measured at low $z$ and the two first CMB peaks;  Godlowski et al. \cite{GS04}, the three first CMB peaks; Bolejko \cite{KB05} uses observed cosmological properties to constraint his models. The main challenge remains to fit the whole set of available cosmological data with a given model. Since very few exact solutions to Einstein's equations can be of use in a cosmological framework, only oversimple toy models have been studied up to now to deal with this issue. However, a project to begin implementing this is currently underway \cite{LH06}.

\section{Conclusion and prospects}

One can find in the literature inhomogeneous cosmological toy models able to solve both the cosmological constant and coincidence problems, i. e., to mimic an ``accelerated expansion'' with no need for dark energy up to the epoch when structure formation enters the nonlinear regime, i. e., around $z \sim 1$.

It has been shown that inhomogeneities likely to solve these problems must be of the subhorizon and strong type, which cannot be studied with perturbation methods. Averaging and smoothing procedures have been proposed which can provide some insight into the issue for very simple cases but which much be used with care since they are not devoided of pitfalls and incompleteness.

Exact solutions of Einstein's equations have the nice property of being able to modelize both strong and weak inhomogeneities. What these models must reproduce is {\it not an accelerated expansion}, which is an artefact of the homogeneous assumption, but the observed dimming of the SN Ia luminosity, i. e., the LDR relation. Some classes have been nicely fitted to this relation at low redshifts. However, the proposed toy models do not pretend to be fair representations of our neighbouring patch of universe. Actually, some of them exhibit controversed properties issued from their spherical symmetry (LTB models) or their matter content (Stephani models). But since very few exact solutions to Einstein's equations can be of use in a cosmological framework, these simple models are usefull to obtain some insight into the issue.

Even if some of them have been shown to match other cosmological observational constraints, the main challenge remains to fit the whole set of available cosmological data. A project to begin implementing this issue is currently underway \cite{LH06}. A solution, if any, would be to use non pathological exact inhomogeneous solutions, reproducing the nearby Universe, coupled to or asymptotically matching nearly homogeneous ones valid up to last scattering.

\section{Acknowledgements} I wish to thank Brandon Carter and Julien Larena for a usefull discussion about averaging procedures. I also want to thank the organizers of the 11th Marcel Grossmann Meeting and specially Andrzej Krasi\'nski to have invited me to present this contribution.

\end{document}